\documentclass[aps,prd,twocolumn,showpacs,superscriptaddress,nofootinbib]{revtex4-1}

\usepackage{amsmath,latexsym}

\usepackage{amssymb}

\begin{document}

\title{Dynamical equivalence of $f(R)$ gravity in Jordan and Einstein frames}

\author{Saikat Chakraborty}
\email{snilch@iitk.ac.in}
\author{Sanchari Pal}
\email{pals@iitk.ac.in}
\affiliation{Department of Physics, Indian Institute of Technology, Kanpur, India}
\author{Alberto Saa} 
\email{asaa@ime.unicamp.br}
\affiliation{
Department of Applied Mathematics, 
 University of Campinas,  13083-859 Campinas, SP, Brazil.}

\date{\today}

\draft

\begin{abstract}
We investigate the dynamics of $f(R)$ gravity in  Jordan and Einstein frames. First, we 
perform a phase-space singularities analysis in both frames. We show that, typically, anisotropic singularities are absent in the Einstein frame, whereas they may appear in the Jordan frame. We conciliate this apparent inconsistency by showing that the necessary conditions  for the existence of the Einstein frame are namely the same ones assuring the absence of the anisotropic singularities in the Jordan frame. In other
words, we show that, at least in the context of Bianchi I cosmologies, the Einstein frame is available only when the original formulation in the Jordan frame is free of anisotropic singularities. 
Furthermore, we present a novel dynamical system formulation  for anisotropic cosmologies in which both
frames, provided they exist,  will be manifestly  equivalent from the dynamical point of view, even though they fail to be diffeomorphic in general.  
Our results could help not only the construction of viable (free of anisotropic singularities) $f(R)$  cosmological models, but also contribute  to the still active debate on the physical interpretation of the two frames.  
\end{abstract}

\maketitle

\section{Introduction} 
\label{Introduction}
Alternative theories of gravity of the $f(R)$ type have been intensively investigated in the last years in
connection with several applications to cosmology and astrophysics, see  \cite{RMP,LR,Rev1,Rev2} for comprehensive
reviews. In all these studies, the dynamical analysis of the modified field equations plays a prominent role, since many important issues as, for instance,   cosmological histories and stability questions of certain astrophysical solutions, are directly related to dynamical properties of the underlying model, see \cite{D1,D2,D3,D4} for some very recent analysis of this kind.  Here, we are concerned with the dynamical
analysis of homogeneous
but anisotropic cosmological models in the  standard $f(R)$ theory of gravity, which can be described in the so-called Jordan frame  by the action 
\begin{equation}
\label{action}
S_J = \frac{1}{2\kappa}\int d^4x \sqrt{-g} f(R) 
+S_M  ,
\end{equation}
where $\kappa = 8\pi G$,    $c=\hbar =1 $, and $S_M$ stands for the usual matter contributions to the total
action, which, in the present case, will be an anisotropic barotropic fluid. By varying (\ref{action}) with respect to the metric, one 
gets
\begin{equation}
\label{eq1}
\left(g_{ab}\Box -\nabla_a\nabla_b \right)F(R) + F(R)R_{ab} - \frac{1}{2}f(R)g_{ab} =
\kappa T_{ab},
\end{equation}
where $F(R)\equiv f'(R)$ and
\begin{equation}
\label{Tab}
T_{ab} = -\frac{2}{\sqrt{-g}} \frac{\delta S_M}{\delta g^{ab}}.
\end{equation}
Provided that $F'=f''\ne 0$, the Euler-Lagrange equations (\ref{eq1}) are fully equivalent to those ones
obtained from the action 
\begin{equation}
\label{action-MS}
S_J = \frac{1}{2\kappa}\int d^4x \sqrt{-g} \left[ F(\varphi) R- V_J(\varphi)   \right] 
+S_M 
\end{equation}
by considering variation with respect to the metric and to the scalar field $\varphi$, where \cite{MagnanoSokolowski}
\begin{equation}
\label{pot}
V_J(\varphi) = \varphi F(\varphi) - f(\varphi).
\end{equation}
Notice that the action (\ref{action-MS})   corresponds to the non-minimally coupled scalar field case considered previously in \cite{ABGS}, but without the  kinetic term of the scalar field, which indeed
prevents their dynamical contents to be equivalent. We will assume hereafter that $F'\ne 0$,
which implies from (\ref{action-MS}) that $\varphi=R$ on the dynamical level. Actions such that $F'>0$ 
are called also $R$-regular in the literature \cite{MagnanoSokolowski}. We will discuss the physical interpretation of this requirement below. 

The action (\ref{action-MS}) can be cast in the so-called Einstein frame by performing the conformal transformation 
\begin{equation}
\label{conf}
\tilde{g}_{ab}  =  {F(\varphi)} {g_{ab}} ,
\end{equation}
which requires, by its own definition, that $F(\varphi)> 0$. 
In terms of the new metric $\tilde{g}_{ab}$, the action of $f(R)$ gravity in the Einstein frame reads
\begin{equation}
\label{Einstein}
S_E = \int d^4x \sqrt{-g} \left[ \frac{1}{2\kappa} \tilde{R}- 
\frac{1}{2}\partial_a\phi\partial^a\phi - V_{E}(\phi)
   \right] 
+\tilde{S}_M ,
\end{equation}
where 
\begin{equation}
\label{varphi}
\phi =\sqrt{\frac{3}{2\kappa}} \ln F(\varphi) 
\end{equation}
and
\begin{equation}
\label{VE}
V_E(\phi) = \frac{V_J(\varphi)}{F(\varphi)^2} .
\end{equation}
The tilde in (\ref{Einstein}) indicates that the quantities are calculated with respect to the new metric $\tilde{g}_{ab}$, defined by 
  the conformal transformation (\ref{conf}), under which the energy-momentum tensor (\ref{Tab}) of the 
matter contributions transforms as
\begin{equation}
\tilde{T}_{a}^{\phantom{a}b}  = \frac{T_{a}^{\phantom{a}b} }{F^2}\,. 
\label{TTtilde}
\end{equation}
Notice that the assumption of a $R$-regular action assures that (\ref{varphi}) is invertible and, hence, that (\ref{VE}) and (\ref{TTtilde}) are, in principle, well defined. Situations in which $F'$ changes
its sign can be considerably more intricate. 
 For example, in case of polynomial $f(R)$ theories with degree of the polynomial higher than two, one in general gets multiple Einstein frame descriptions, leading to various possible ambiguities \cite{Bhattacharya:2015nda}. 
 The condition $F'<0$, which would also guarantee the invertibility of 
(\ref{varphi}), 
  is related to unbounded growth of curvature perturbations in the presence of matter \cite{Dolgov:2003px,Faraoni:2006sy}, so the condition  $F'>0$  ($R$-regularity) is usually imposed in $f(R)$
   gravity. As for the
  condition  $F>0$, 
it is also sometimes argued that the effective gravitational coupling in $f(R)$ gravity is played by the quantity $\kappa/F$, whose positivity requires $F>0$. 
 However, it is indeed possible to have regular  isotropic cosmological solutions   crossing regions where
 $F=0$, see \cite{GSBFRF,ABG} for instance.

 On the other hand, phase-space regions where $F=0$
are known to be associate with severe and dynamically unavoidable anisotropic singularities in the
Jordan frame, see  
\cite{ABGS,Figueiro:2009mm,e3}. As we will see, such singularities are, by construction, absent in the Einstein frame, which could indicate a physical incompatibility between the two descriptions of the
same theory. We figure out this apparent inconsistency  by showing that the 
necessary conditions for the existence of the Einstein
frame, namely $F>0$ and $F'>0$, are sufficient for  assuring the absence of the anisotropic singularities in the Jordan
frame. Moreover, we will show that, when both frames exist, they are fully equivalent from
the dynamical point of view, even though the dynamical system formulation in the two frames   are not in general   diffeomorphic or, in the dynamical
system language, topologically equivalent.
Such results could help not only the construction of viable   $f(R)$ cosmological models, but also contribute  to the still active debate
on the physical interpretation of the two frames, see \cite{e1,e2,e20,e21,e22,e3,e4,e5,e500,e501,e51,e52} for other recent references on this issue and many of its implications in different physical contexts. In particular, for the question on  the frame equivalence at the quantum level, see \cite{e61,e62,e63}.

 In the next section, we will derive all the pertinent
equations, in the Jordan and Einstein frames, for a Bianchi-I homogeneous but anisotropic universe filled with
an anisotropic barotropic fluid. Phase-spaces for both frames are determined and we will   prove that
the existence of the Einstein frame rests effectively on the conditions $F>0$ and $F'>0$, which, on the other hand, guarantee that the Jordan frame description is free from anisotropic singularities. In Section III, we present a novel dynamical formulation for anisotropic
cosmologies in both frames. The new dynamical variables allow a one-to-one correspondence between all dynamical quantities in both frames, establishing their dynamical equivalence. We notice that the 
dynamical system approach for homogeneous and isotropic Friedmann-Lemaitre-Robertson-Walker (FLRW)  spacetime filled with a perfect fluid in generic $f(R)$ gravity       has traditionally been formulated in terms of expansion normalized dynamical variables,   see \cite{D2,Amendola:2006we, Carloni:2007br, Carloni:2004kp}, although it is also possible to formulate in alternative ways, see, for instance, \cite{D1, Carloni:2015jla}. The dynamics of metric shear for $R^n$ gravity in vacuum and in presence of an isotropic fluid have previously been also studied using the $(1+3)$ covariant formalism, see \cite{Leach:2006br,Goheer:2007wu}. However, a more general dynamical system formulation involving an anisotropic fluid was still missing. Our new approach  allows us to establish the
classical equivalence between the two frames in a new way 
which has not been attempted before.
The last section is devoted to some concluding remarks.

\section{Bianchi-I cosmological dynamics}

We will focus on the dynamics of a generic $f(R)$ theory of gravity with an anisotropic fluid
in both Jordan (Eqs. (\ref{action}) or (\ref{action-MS})) and Einstein frames (Eq. (\ref{Einstein})).  
 The homogeneous and anisotropic Bianchi-I metric 
\begin{equation}
ds^2 = -dt^2 + a^2_1(t)dx_1^2+ a^2_2(t)dx_2^2+ a^2_3(t)dx_3^2
\end{equation}
is the simplest situation where we can explore properly the differences between the two frames.  
By introducing the   average expansion factor $a (t) = \sqrt[3]{a_1a_2 a_3}$, we 
can parameterize    the Bianchi-I metric  as
\begin{equation}
ds^2=-dt^2+a^2(t)\left[e^{2\beta_1(t)}dx_1^2+e^{2\beta_2(t)}dx_2^2+e^{2\beta_3(t)}dx_3^2\right],
\label{bianchi-I}
\end{equation}
where, by construction,  
\begin{equation}
\label{bianchi_constraint_2}
\beta_1 +  \beta_2 +  \beta_3 = 0.
\end{equation}
For further references on this parametrization, see \cite{t1,Chakraborty:2018thg}.
The metric (\ref{bianchi-I}) has three independent dynamical variables in $f(R)$ gravity, and we choose them to be
\begin{equation}
\label{Hb}
H = \frac{\dot a}{a},\quad  \dot\beta_\pm = \dot\beta_1 \pm \dot\beta_2.
\end{equation} 
The quantity
\begin{equation}
 \sigma^2=\dot{\beta}_1^2+\dot{\beta}_2^2+\dot{\beta}_3^2 = \frac{3}{2}\dot{\beta}_+^2 + \frac{1}{2}\dot{\beta}_-^2
\end{equation}
measures the total amount of anisotropy of (\ref{bianchi-I}). 
 Observe that when   $\sigma=0$, the spatial coordinates can be suitably rescaled to recast (\ref{bianchi-I}) in the standard spatially flat  FLRW  form. The Ricci scalar
 for the metric (\ref{bianchi-I}) reads simply
 \begin{equation}
 \label{scalar}
 R = 6\dot H + 12H^2 + \sigma^2.
 \end{equation}

We will assume that the universe described by  (\ref{bianchi-I}) is filled with an 
 anisotropic barotropic  fluid with 
   energy momentum tensor parametrized as \cite{EoS}  
\begin{equation}
\label{baro}
T_{a}^{\phantom{a}b} = {\rm diag}\left( -\rho, p_1, p_2 , p_3  \right) =
 {\rm diag}\left( -\rho, \omega_1\rho,  \omega_2\rho ,  \omega_3\rho  \right).
\end{equation}
We define the anisotropic    equation of state  as
\begin{equation}
p_i = (\omega +\delta_i)\rho,
\end{equation}
with $i=1,2,3$, where $\omega$ is the average barotropic parameter and 
$
 \omega_i=\omega+\delta_i,$
with $\delta_1+\delta_2+\delta_3=0$ by construction. We will parameterize our fluid by the constants $\omega$ and
$\delta_\pm = \delta_1 \pm \delta_2$, whose meaning are rather clear.

\subsection{The Jordan frame}

For sake of completeness, we will write the $f(R)$ equations in both frames also for the isotropic fluid case. However, from the dynamical point of view, our primary interest is the phase-space anisotropic singularities, which appear only for the anisotropic fluids. Of course, once we establish the dynamical
equivalence of both frames for anisotropic fluids, the isotropic case follows naturally as a simple corollary. 

\subsubsection{Isotropic fluid}

 The dynamics of the Bianchi-I  metric (\ref{bianchi-I}) under $f(R)$ gravity in presence of such an isotropic fluid can be described by the following set of equations \cite{Bhattacharya:2017cbn,Chakraborty:2018thg},
\begin{equation}
3H^2=\frac{\kappa}{F}\left(\rho+\frac{RF-f}{2\kappa}-\frac{3HF'\dot{R}}{\kappa}\right)+\frac{1}{2}\sigma^2,
\label{bianchi_constraint_1}
\end{equation}
\begin{eqnarray}
2\dot{H}+3H^2&=&-\frac{\kappa}{F}\left(\omega\rho+\frac{\dot{R}^2F''+(2H\dot{R}  +\ddot{R})F'}{\kappa}\right. \nonumber \\
&& \left.\quad\quad\quad\quad\quad -\frac{RF-f}{2\kappa}\right)  
 -\frac{1}{2}\sigma^2, 
\label{bianchi_dynamic_1}
\end{eqnarray}
\begin{eqnarray}
 \dot{\rho}+3H\left(1+\omega\right)\rho&=&0,
\label{bianchi_dynamic_2}\\
  \ddot{\beta_i}+\left(3H+\frac{\dot{R}F'}{F}\right)\dot{\beta_i}&=&0,
  \label{bianchi_dynamic_3}
\end{eqnarray}
where $i=1, 2, 3$.
It is easy to check that  $\sigma$ 
obeys the a dynamical equation of the same form as obeyed by the $\dot{\beta}_i$ themselves
\begin{equation}
\dot{\sigma}+\left(3H+\frac{\dot{R}F'}{F}\right)\sigma=0.
\label{bianchi_dynamic_345}
\end{equation}
If we know the form of the function $f(R)$ and concentrate only on the quantity $\sigma(t)$, we see that there are now a total of three functions of time $H(t),\,\rho(t),\,\sigma(t)$ governing the dynamics. The existence of the constraint equation (\ref{bianchi_constraint_1}) implies that only two of them are independent. Without loss of generality, we can choose them to be $H(t)$ and $\sigma(t)$. Given some form of the function $f(R)$, they can be determined by solving equations (\ref{bianchi_dynamic_1}) and (\ref{bianchi_dynamic_345}), and then   $\rho(t)$ can then be found using the constraint equations (\ref{bianchi_constraint_1}). Notice that, despite  involving only the two dynamical variables $H(t)$ and $\sigma(t)$, the underlying phase-space is higher dimensional, since equation (\ref{bianchi_dynamic_1}) involves
the third derivative of $H$. We will return to this point in the next section. 

\subsubsection{Anisotropic fluid}

As for the case of an anisotropic fluid, the dynamics of 
  the Bianchi-I metric  (\ref{bianchi-I}) under $f(R)$ is governed by a set of equations analogous to
 (\ref{bianchi_constraint_1}) - (\ref{bianchi_dynamic_3}), but now with
  \begin{eqnarray}
 &\displaystyle \dot{\rho}+ \left( 3H\left(1+\omega\right) +  \boldsymbol{\delta}\cdot  \dot{\boldsymbol{\beta}} \right)\rho =0 , &
\label{bianchi_dynamic_2_anifl} \\
&  \displaystyle \ddot{\beta_i}+\left(3H+\frac{\dot{R}F'}{F}\right)\dot{\beta_i}=\frac{\kappa\rho}{F}\delta_i,& \label{bianchi_dynamic_3_anifl} 
  \end{eqnarray}
where  $i=1, 2, 3,$ and
 \begin{equation}
 \label{bolddelta}
\boldsymbol{\delta}\cdot  \dot{\boldsymbol{\beta}} = \delta_1\dot{\beta_1}+\delta_2\dot{\beta_2}
+\delta_3\dot{\beta_3} = \frac{3}{2}\delta_+\dot{\beta}_+  + \frac{1}{2}\delta_-\dot{\beta}_-,
 \end{equation}
instead of (\ref{bianchi_dynamic_2}) and (\ref{bianchi_dynamic_3}).
Observe that  the quantity $\sigma$ 
in this case does not obey a simple equation like Eq.~(\ref{bianchi_dynamic_345}), the reason being that the right hand sides of the three $\beta$-equations (\ref{bianchi_dynamic_3_anifl}) can be all different.
In this case, knowing the form of the function $f(R)$, we cannot proceed concentrating only on the quantity $\sigma(t)$. Instead, we see that there are in this case a total of five functions of time $H(t),\,\rho(t),\,\beta_1(t),\,\beta_2(t),\,\beta_3(t)$ governing the dynamics. The existence of the constraint equations (\ref{bianchi_constraint_2}) and (\ref{bianchi_constraint_1}) implies that only three of them are independent, and we choose them as $H(t)$ and $\beta_{\pm}(t)$, see (\ref{Hb}).  Given some form of the function $f(R)$, $H(t)$ and $\beta_{\pm}(t)$ can be determined by solving Eq.~(\ref{bianchi_dynamic_1}) and (\ref{bianchi_dynamic_3_anifl}). As for the isotropic case, 
 $\rho(t)$ can then be found using the constraint equation  (\ref{bianchi_constraint_1}). 

\subsection{The Einstein frame}

The construction of the Einstein frame    is based on the conformal transformation (\ref{conf}).  One can also recast the conformally related Einstein frame metric
$\tilde{g}_{\mu\nu}$ in a Bianchi-I form
\begin{equation}
d\tilde{s}^2=Fds^2 \label{bianchi-I_tilde} 
=  -d\tilde{t}^2 + \tilde{a}^2 \left[e^{2\tilde{\beta}_1}dx_1^2+e^{2\tilde{\beta}_2}dx_2^2+e^{2\tilde{\beta}_3}dx_3^2\right] ,
\end{equation}
by using the redefinitions
\begin{eqnarray}
d\tilde{t}&=&\sqrt{F}dt,\\
\tilde{a}\left(\tilde{t}\right)&=&\sqrt{F}a(t), \\
\tilde{\beta}_i\left(\tilde{t}\right)&=&\beta_i(t) ,
\label{metric_redef}
\end{eqnarray}
with $i=1,2,3$. The 
Hubble parameter in the Einstein frame is defined as
\begin{equation}
\tilde{H}=\frac{\tilde{a}^{\prime}}{\tilde{a}}
\end{equation}
where the prime denotes here the derivative with respect to the Einstein frame time variable $\tilde{t}$. The Hubble parameters of the two frames are known to be related as \cite{Paul:2014cxa}
\begin{equation}
H=\sqrt{F}\left(\tilde{H}-\sqrt{\frac{\kappa}{6}}\,\phi'\right)\,,
\label{hhtilde}
\end{equation}
which for our purposes can be better expressed as
\begin{equation}
\tilde{H}=\frac{H}{\sqrt{F}}\left(1+\frac{\dot{R}F'}{2 HF}\right).
\label{htildeh}
\end{equation}
From (\ref{TTtilde}) and (\ref{baro}), one can write
\begin{eqnarray}
\tilde{\rho}=\frac{\rho}{F^2}, \quad \tilde{p}_i=\frac{p_i}{F^2},
\label{rhorhotilde_pptilde_anifl} 
\end{eqnarray}
for an anisotropic fluid. Note that the barotropic equations of state remains the same in both  frames.
It is also straightforward to check that the total amount of anisotropy $\tilde{\sigma}$ in the Einstein frame defined as
\begin{equation}
\tilde{\sigma}^2=\tilde{\beta}_1^{\prime 2}+\tilde{\beta}_2^{\prime 2}+\tilde{\beta}_3^{\prime 2}
\end{equation}
is related to total amount of anisotropy $\sigma$ in the Jordan frame as
\begin{equation}
\tilde{\sigma}^2=\frac{\sigma^2}{F},
\label{sigmasigmatilde}
\end{equation}
from where one can envisage the possibility of having some problems for anisotropic solutions when $F=0$. We will return to this point below. 

\subsubsection{Isotropic fluid}
The dynamics of the Bianchi-I metric (\ref{bianchi-I_tilde}) in the 
  Einstein frame formulation of $f(R)$ gravity given by the action (\ref{Einstein}), with
  an isotropic fluid,   is governed by the following set of equations
\begin{eqnarray}
&\displaystyle 3\tilde{H}^2 = \kappa\left(\frac{1}{2}\phi^{\prime 2}+V(\phi)+\tilde{\rho}\right)+\frac{\tilde{\sigma}^2}{2},&
\label{bianchi_constraint_1_tilde}\\
&\displaystyle 2\tilde{H}^{\prime}+3\tilde{H}^2 =-\kappa\left(\frac{1}{2}\phi^{\prime 2}-V(\phi)+\omega\tilde{\rho}\right)-\frac{\tilde{\sigma}^2}{2} ,&
\label{bianchi_dynamic_1_tilde}\\
&\displaystyle \tilde{\rho}' + \left(\sqrt{\frac{\kappa}{6}}(1-3\omega) \phi'
+3\tilde{H}(1+\omega)\right)\tilde{\rho}=0,&
\label{bianchi_dynamic_2_tilde}\\
&\displaystyle \tilde{\sigma}'+3\tilde{H}\tilde{\sigma}=0,&
\label{bianchi_dynamic_345_tilde}\\
&\displaystyle \phi''+ 3\tilde{H} \phi'+\frac{dV}{d\phi}=\sqrt{\frac{\kappa}{6}}(1-3\omega)\tilde{\rho} .&
\label{bianchi_dynamic_6_tilde}
\end{eqnarray}
We see that there are now a total of four functions of time $\tilde{H}(\tilde{t}),\,\tilde{\rho}(\tilde{t}),\,\tilde{\sigma}(\tilde{t}),\,\phi(\tilde{t})$ governing the dynamics in Einstein frame, whereas there was only three functions of time governing the Jordan frame dynamics. The existence of the constraint equation (\ref{bianchi_constraint_1_tilde}) implies that only three of them are independent, as opposed to two in Jordan frame. Without loss of generality, we can choose them to be $\tilde{H}(\tilde{t})$, $\phi(\tilde{t})$ and $\tilde{\sigma}(\tilde{t})$. Given some form of the function $f(R)$, we can find the Einstein frame scalar field potential $V(\phi)$ by using Eq.~(\ref{varphi}) and Eq.~(\ref{VE}). We can then solve Eq.~(\ref{bianchi_dynamic_1_tilde}), Eq.~(\ref{bianchi_dynamic_6_tilde}) and Eq.~(\ref{bianchi_dynamic_345_tilde}) to get $\tilde{H}(\tilde{t})$, $\phi(\tilde{t})$ and $\tilde{\sigma}(\tilde{t})$. The fluid energy density $\tilde{\rho}(\tilde{t})$ can then be determined from Eq.~(\ref{bianchi_constraint_1_tilde}).

\subsubsection{Anisotropic fluid}
Since for an anisotropic fluid the individual barotropic constants $\omega_i$  remain the same in both frames, the average barotropic constant $\omega$ and the deviations $\delta_i$  from it also remain unaltered. Moreover, we have  $\tilde{\beta}_{\pm}(\tilde{t})=\beta_{\pm}(t)$ and
\begin{equation}
\tilde{\beta}_{\pm}^{\prime 2}=\frac{\dot{\beta}_{\pm}^2}{F}.
\label{betadotbetaprimetilde_anifl}
\end{equation}
The dynamics of $f(R)$ gravity in the Einstein frame  with an anisotropic fluid is  governed by a set of equations analogous to (\ref{bianchi_constraint_1_tilde}) - (\ref{bianchi_dynamic_6_tilde}), but
now with 
\begin{eqnarray}
& \displaystyle \tilde{\rho}' + \left(\sqrt{\frac{\kappa}{6}}(1-3\omega) \phi'+3\tilde{H} (1+\omega) +  \boldsymbol{\delta}\cdot   {\boldsymbol{\beta}}'\right)\tilde\rho = 0\,,&
\label{bianchi_dynamic_2_tilde_anifl} \\
&  
\tilde{\beta}_{i}''+3\tilde{H}\tilde{\beta}_{i}'=\kappa\tilde{\rho}\delta_{i}\,,&
\label{bianchi_dynamic_345_tilde_anifl}
\end{eqnarray}
where $ \boldsymbol{\delta}\cdot   {\boldsymbol{\beta}}'$ is defined analogously to (\ref{bolddelta}),
instead of (\ref{bianchi_dynamic_2_tilde}) and (\ref{bianchi_dynamic_345_tilde}).
Again, given some form of the function $f(R)$, we can find the Einstein frame scalar field potential $V(\phi)$ by using Eq.~(\ref{varphi}), Eq.~(\ref{VE}). We can then solve Eq.~(\ref{bianchi_dynamic_1_tilde}), Eq.~(\ref{bianchi_dynamic_6_tilde}), and  Eq.~(\ref{bianchi_dynamic_345_tilde_anifl}) to get $\tilde{H}(\tilde{t})$, $\phi(\tilde{t})$ and $\tilde{\beta}_{\pm}(\tilde{t})$. The energy density of the anisotropic fluid $\tilde{\rho}$ can then be determined from Eq.~(\ref{bianchi_constraint_1_tilde}).

\subsection{The $F=0$ submanifold}

The conformally related Einstein frame description becomes invalid on the phase-space submanifold $f'(R)\equiv F(R)=0$, because the very definition of the conformal transformation (\ref{conf}) does not hold there. Typically, the Einstein frame scalar field (\ref{varphi}) and its potential (\ref{VE}) becomes singular at the submanifold
given by $F=0$. Nevertheless, it is indeed possible to have regular homogeneous and isotropic cosmological solutions crossing such submanifold, see for some explicit examples \cite{GSBFRF,ABG}. However, the situation
is qualitatively different for the case of anisotropic solutions. 
As we have mentioned in section \ref{Introduction}, for anisotropic cosmologies, the phase-space submanifold $F =0$ does not only imply a mathematical difficulty, but is an actual physical singularity which,
typically, is dynamically unavoidable, challenging the physical viability of the underlying
model. This issue was first discussed in \cite{ABGS} for nonmimally coupled $f(\varphi)R$ gravity theories, and then was generalized in \cite{Figueiro:2009mm} for a more general $f(R,\varphi,X)$ gravity theories, where $X=-\frac{1}{2}\partial_{\mu}\phi\partial^{\mu}\phi$ is the canonical kinetic term of the scalar field.  Although in both the papers the authors do not assume the presence of any hydrodynamic fluid, the conclusion remains the same even if we add one. See also \cite{e3} for other instances of the singularities on the $F=0$ submanifold. In what follows, we show how these physical singularities also arise in the Jordan frame of $f(R)$ gravity theories with anisotropic fluids. 

Let us introduce the Hubble parameter $H_i$ associated with the $i$-direction of the metric (\ref{bianchi-I}), {\em i.e},
\begin{equation}
H_i= \frac{\dot a_i}{a_i} = H+\dot{\beta}_i.
\end{equation} 
The Kretschman scalar $I=R_{abcd}R^{abcd}$ for the Bianchi-I metric is given by
\begin{equation}
\frac{1}{4}I=\sum_{i=1}^3\left(\dot{H}_i+H_i^2\right)^2 + H_1^2H_2^2 + H_2^2H_3^2 + H_3^2H_1^2.
\end{equation}
It is clear that any 
divergence of $H_i$  will imply a divergence of the Kretschman scalar, meaning  a real spacetime singularity.
From equations   (\ref{bianchi_dynamic_3_anifl}), we have
\begin{equation}
\label{div}
  F\ddot{\beta}_\pm =  -\left(3HF+ {F'\dot{R}} \right)\dot{\beta}_\pm + \kappa\rho   \delta_\pm.
\end{equation}
Notice that, from (\ref{bianchi_constraint_1}), we have on $F=0$
\begin{equation}
F'\dot{R} = \frac{2\kappa  \rho    - f}{6H}  .
\end{equation}
It is clear that (\ref{div}) is ill-defined on the submanifold $F=0$, where the phase-space flow
has a singularity which implies divergence of $\dot{\beta}_\pm$ and, consequently, the divergence of
the Kretschman scalar $I$. Note the important role played by metric anisotropy here. The physical singularity arises because the metric anisotropy $\sigma$ diverges. This kind of singularity on the $F=0$ submanifold can be avoided in homogeneous and isotropic spacetimes, although the mathematical difficulty to define an Einstein frame  in such cases still remains. 
 
On the other hand, 
 one might say, naively, that there are no such physical singularities in the Einstein frame, 
since in this case the  theory is essentially general relativity where this kind of an anisotropic singularity cannot arise. Indeed, no singularity is apparent from the equations   (\ref{bianchi_dynamic_345_tilde_anifl}). This might cast some doubts on the physical equivalence of the two frames at the classical level. This naive conclusion, however, is incorrect, as the Einstein frame description is itself broken if we have $F=0$ somewhere in the phase-space. The very conditions for the existence and uniqueness of an Einstein frame are the same ones which assure the absence of   anisotropic singularities in the Jordan frame. Therefore, the issue of conformal inequivalence does not arise here. 
From now on, we assume that a well defined $f(R)$ theory of gravity requires $F>0$ and $F'>0$. If this cannot be guaranteed for all the phase-space, then our subsequent consideration applies only in the domain where these conditions are met.

\section{Dynamical system analysis}

Provided the requirements for the existence and uniqueness  of the Einstein frame hold, namely $F>0$ and $F'>0$, we can compare the dynamics of anisotropic cosmologies in $f(R)$ gravity in both the Jordan and Einstein frames.
In this section, we will introduce a novel formulation for the dynamical variables  which
will allow us to introduce an one-to-one correspondence between all dynamical quantities
in both frames, establishing their    complete equivalence from the  dynamical point of view.  

\subsection{Jordan frame}

As in the last section, for sake of completeness and comparison with previous works, we will also present the explicit results for
the isotropic case. 

\subsubsection{Isotropic fluid}

In presence of an isotropic fluid, one can rewrite the Hamiltonian constraint equation  (\ref{bianchi_constraint_1}) in the Jordan frame as
\begin{equation}
\label{eqConstr}
3FH^{2}=\frac{RF-f}{2}-3H\dot{R}F'+\kappa\rho+\frac{ \sigma^2F}{2}.
\end{equation}
Notice that each term  has a mass dimension $[M]^2$. Let us now introduce our dimensionless dynamical quantities by multiplying each term by $\frac{\kappa}{F^2}$. Since $F$ is a dimensionless quantity by itself,   it will have no effect on the dimension (or, better to say, the dimensionlessness) of the dynamical variables; it has been introduced only for future convenience. Our set of dimensionless dynamical variables are as follows
\begin{eqnarray}
U_{1}&=&\frac{3\kappa H^{2}}{F}, \\
U_{2}&=&\frac{\kappa(RF-f)}{2F^2},\label{U2} \\
U_{3}&=&\frac{3\kappa H\dot{R}F'}{F^2},\label{U3}\\
U_{4}&=&\frac{\kappa \sigma^{2}}{2F}, \label{U4} \\
U_{5}&=&\frac{\kappa^{2}\rho}{F^2}.
\end{eqnarray}
The Hamiltonian constraint then reads
\begin{equation}
 U_{1}-U_{2}+U_{3}-U_{4} = U_5
\label{phase_space_const}.
\end{equation}
Let us substitute in place of the cosmological time $t$ the monotonically increasing dimensionless variable
$N$ corresponding to
\begin{equation}
{\epsilon}N= {\ln a},
\end{equation}
where $\epsilon$ is defined to be $+1$ for expanding universe and $-1$ for a contracting one.
The variable $N$ is called the logarithmic time, and we choose the scale factor at $t=0$ to be $a_0=1$. Therefore as time progresses in the forward (positive) direction, $N$ becomes positive and goes towards $+\infty$ in case of both the expanding and contracting universes. We see that
\begin{eqnarray}
{\epsilon}\dot{N}= {H},
\end{eqnarray}
so that $\dot{N}$ is always positive, {\em i.e.} $N$ is always monotonically increasing with time. This justifies taking $N$ as the time variable in both expanding and contracting universe. Around a bounce or a turnaround point, this will not be valid though.

There are five dynamical variables and, hence, 
the underlying phase-space is five-dimensional. Since there is 
 a constraint, effectively there are only four independent dynamical variables. Without loss of generality we can take $U_{1}$, $U_{2}$, $U_{3}$ and $U_{4}$ as our independent variables, and  $U_5$ can be determined from the constraint equation Eq.~(\ref{phase_space_const}). The corresponding dynamical equations are found out by taking the derivative of the dynamical variables with respect to the dimensionless time variable $N$  and comparing with the equations (\ref{bianchi_constraint_1}) - (\ref{bianchi_dynamic_3}). They are:
\begin{eqnarray}
\frac{dU_{1}}{dN}&=&\epsilon[-4U_{1}+2U_{2}-U_{3}-2U_{4}+\gamma(U_2)], \label{dU1}\\
\frac{dU_{2}}{dN}&=&\epsilon\frac{U_{3}}{2U_{1}}\left[-2U_{2}+\gamma(U_2)\right], \label{dU2}\\
\frac{dU_{3}}{dN}&=&\epsilon\bigg[(1-3\omega)U_1+(1+3\omega)U_2-(4+3\omega)U_3  \nonumber \\
&& -(1-3\omega)U_4+\frac{U_3}{U_1}(U_2-2U_3-U_4)\nonumber \\
&& \left. +\gamma(U_2)\left(\frac{U_3}{2U_1}-1\right)\right],\label{dU3}\\
\frac{dU_{4}}{dN}&=&-6\epsilon U_{4}\left(1+\frac{U_{3}}{2U_{1}}\right). \label{dU4}
\end{eqnarray}
where 
\begin{equation}
\label{gamma}
\gamma(U_2)=\frac{\kappa f}{F^2}.
\end{equation} 
The function $\gamma(U_2)$ is defined as follows. 
Notice that, by construction, $U_2$ is a function of $R$ only and it could be, in principle, inverted to find $R(U_2)$. Since $ \frac{\kappa f}{F^2}$ is a function of $R$  and, therefore, a function of $U_2$ only,   we denote it by $\gamma(U_2)$. We will return to the issue of the invertibility of $U_2$  in the last section. 

\subsubsection{Anistropic fluid}

In the presence of an anisotropic fluid, one cannot use $\sigma$ as a dynamical variable anymore. Instead of
$U_4$ given by (\ref{U4}), we need to introduce two new variables
\begin{eqnarray}
U_{4}^+&=&\frac{3}{4}\frac{\kappa\dot{\beta}_+^{2}}{F}, \\
U_{4}^-&=&\frac{1}{4}\frac{\kappa\dot{\beta}_-^{2}}{F},
\end{eqnarray}
in terms of which the Hamiltonian constraint reads
\begin{equation}
 U_{1}-U_{2}+U_{3}-U_{4}^+-U_{4}^- = 
U_{5} .
\label{phase_space_const_anifl}
\end{equation}
There are now six dynamical variables and one constraint equation and so, effectively, there are five independent dynamical variables. Without loss of generality we can take $U_{1}$, $U_{2}$, $U_{3}$ and $U_{4}^{\pm}$ as   independent variables, and $U_5$ can be determined from the constraint equation Eq.~(\ref{phase_space_const_anifl}).
The dynamical equations for the anisotropic case are the same Eq. (\ref{dU1}), (\ref{dU2}), and (\ref{dU3}), remembering that $U_4= U_{4}^++U_{4}^-$, and the new pair of equations
\begin{eqnarray}
\frac{dU_{4}^+}{dN}&=&\epsilon\left[-6U_4^+\left(1+\frac{U_3}{2U_1}\right)+3\delta_+U_5\sqrt{\frac{U_4^+}{U_1}}\right],\\
\frac{dU_{4}^-}{dN}&=&\epsilon\left[-6U_4^-\left(1+\frac{U_3}{2U_1}\right)+\sqrt{3}\delta_-U_5\sqrt{\frac{U_4^-}{U_1}}\right],
\end{eqnarray}
instead of (\ref{dU4}).

\subsection{Einstein frame}

We will proceed for the Einstein frame in the same way we did for the Jordan case. In particular, we will
also introduce a logarithmic time variable $\tilde N$
\begin{equation}
{\tilde{\epsilon}}\tilde{N}= {\ln\tilde{a}},
\end{equation}
where $\tilde{\epsilon}$ is defined to be $+1$ if the universe is expanding from the Einstein frame point of view ({\em i.e.} $\tilde{a}(\tilde{t})$ is increasing with $\tilde{t}$) and $-1$ if the universe is contracting from the Einstein frame point of view.  
\subsubsection{Isotropic fluid}

In presence of an isotropic fluid, the Hamiltonian constraint equation in the Einstein frame frame is given by Eq.~(\ref{bianchi_constraint_1_tilde}). Let us now define the dimensionless dynamical variables in the Einstein frame as follows
\begin{eqnarray}
\tilde{U}_{1}&=&3\kappa\left(\tilde{H}-\sqrt{\frac{\kappa}{6}} \phi' \right)^{2}, \\
\tilde{U}_{2}&=&\kappa^{2}V(\phi),\label{tildeU2}\\ 
\tilde{U}_{3}&=&\sqrt{6\kappa^{3}} \left(\tilde{H}-\sqrt{\frac{\kappa}{6}}\phi'\right)\phi',\\ 
\tilde{U}_{4}&=&\frac{\kappa}{2}\tilde{\sigma}^{2}, \label{tildeU4}\\
\tilde{U}_{5}&=&\kappa^{2}\tilde{\rho}.
\end{eqnarray}
Notice that we have, by construction,
\begin{equation}
\tilde{U}_2=U_2\,\,,\,\,\tilde{U}_4=U_4\,\,,\,\,\tilde{U}_5=U_5\,\,.
\end{equation}
The
Hubble parameters $H$ and $\tilde H$ in the two frames are related by Eq.~(\ref{htildeh}). This relation, when used in the definition of $\tilde{U}_{1}$, gives back exactly the form of $U_{1}$, {\em i.e.}, we also have effectively $\tilde{U}_1=U_1$. Regarding $\tilde{U}_3$, note that
\begin{equation}
\phi'=\sqrt{\frac{3}{2\kappa}}\frac{d\ln F}{dt}\frac{dt}{d\tilde{t}}=\frac{F'\dot{R}}{{F^\frac{3}{2}} }
\end{equation}
The above relation, along with Eq.~(\ref{htildeh}), when inserted in the definition of $\tilde{U}_{3}$, it also takes the form of $U_{3}$. Therefore we have
$
\tilde{U}_i=U_i,
$ 
for all $i$, 
and here lies the great advantage of constructing the dimensionless dynamical variables in the particular way that we have taken. The dynamical variables in the two frames have a one-to-one correspondence, which
of course implies that both phase-space are diffeomorphic.  Such one-to one-correspondence   considerably reduces our effort in finding out the constraint equation and the dynamical equations in Einstein frame. For example,  the dynamical variables in the Einstein frame   satisfy the same constraint equation as the dynamical variables in the Jordan frame, namely
\begin{equation}
\tilde{U}_{1}-\tilde{U}_{2}+\tilde{U}_{3}-\tilde{U}_{4} = \tilde{U}_{5}.
\end{equation}
For the true dynamical equations (\ref{dU1}) - (\ref{dU4}), we need to take into consideration the change from $N$ to $\tilde N$.  For this purpose, notice that 
\begin{equation}
\label{Om1}
\frac{dN}{d\tilde{N}} = \frac{\dot N}{\tilde N'} \frac{dt}{d\tilde t} = \frac{\tilde{\epsilon}}{\epsilon}\left(\frac{2U_1}{2U_1+U_3}\right) = \Omega(U_1,U_3),
\end{equation}
where (\ref{htildeh}) was used.  
 Now, knowing the Jordan frame dynamical equation 
\begin{equation}
\frac{dU_i}{dN}=f_{i}(U_{1},U_{2},U_{3},U_{4})
\end{equation}
for the Jordan frame dynamical variable $U_i$, the corresponding Einstein frame dynamical equation 
will be 
\begin{equation}
\label{neq1}
\frac{dU_i}{d\tilde N}=\tilde f_{i}(\tilde U_{1},\tilde U_{2},\tilde U_{3},\tilde U_{4})
\end{equation}
with
\begin{equation}
\label{neq2}
\tilde f_{i}(\tilde U_{1},\tilde U_{2},\tilde U_{3},\tilde U_{4}) = 
\Omega(\tilde U_1,\tilde U_3)
f_{i}(\tilde U_{1},\tilde U_{2},\tilde U_{3},\tilde U_{4}).
\end{equation}
 Note the fundamental part played by the one-to-one correspondence property. We could only 
exchange $\tilde U_i$ and $U_i$ in all expressions  precisely because one has $\tilde{U}_i=U_i$ for all $i$.

\subsubsection{Anisotropic fluid}

The situation for the anisotropic fluid in the Einstein frame is analogous to the Jordan case. Since
one cannot use $\sigma$ as dynamical variable anymore, the variable $\tilde U_4$ given by (\ref{tildeU4}) must
be split as
\begin{equation}
\tilde U_4 = \tilde U_4^+ + \tilde U_4^- = 
\frac{3\kappa}{4}\tilde{\beta}_+^{\prime^2} + \frac{\kappa}{4}\tilde{\beta}_-^{\prime^2} ,
\end{equation}
in the same way we have done for the Jordan frame case. The equations for the anisotropic fluid
case are obtained in the same way we did for the isotropic case, by means of (\ref{neq1}) and 
(\ref{neq2}).

\subsection{Dynamical equivalence}

We are now ready to prove one of our central results, the complete dynamical equivalence of both
frames.Two autonomous dynamical systems $\dot x = f_1(x)$, $x\in\mathbb{R}^n$, and  $\dot y = f_2(y)$, $y\in\mathbb{R}^n$, will be dynamically equivalent (or topologically equivalent in the dynamical system language, see, for instance, \cite{Arnold})  if there exists a homeomorphism (diffeomorphism in the present case) $y=h(x)$ which maps  solutions $x(t)$ into solutions $y(t)$ preserving the direction 
of time, meaning that if  $x(t)$  is a solution of the fist dynamical set of equations, $y(t) = h(x(t))$ will be a solution of the second one. The idea behind the concept of topological equivalence for dynamical systems is rather simple: if two systems are topologically equivalent, their dynamical contents are equivalent in the sense that one can map the evolution of any observable in both systems in a one-to-one manner. In particular, all dynamical properties of certain solutions as, for instance, fixed points and their attractive/repulsive nature, periodic solutions, limit cycles, among others, are preserved from one system to the other. 

In our case, the dynamical variables were constructed in order to assure that $\tilde U_i = U_i$, {\em i.e.}, $h$ is the identity map, establishing that the phase space of the two frames are trivially diffeomorphic. However, the dynamical equations in the two frames are not topologically equivalent in general. The condition of mapping solutions into solutions of the type $y(t) = h(x(t))$ implies that the
vector fields of the two dynamical systems obey  $f_2 = \left(\nabla_xh \right)f_1$, which is not observed for our case,
since we have (\ref{neq2}). The dynamical system formulation in the two frames will be topologically equivalent, and consequently also dynamically equivalent, if $\Omega=1$. We will return to this issue below. 

However, the topological equivalence is a stronger than necessary requirement to assure dynamical equivalence in our case. Let us analyze more closely the function $\Omega$. We have
\begin{eqnarray}
\Omega(U_1,U_3)= \frac{1}{\sqrt{F}}\frac{\tilde{\epsilon}}{\epsilon }\frac{H}{\tilde{H}}
\label{Omega}.
\end{eqnarray}
Since $F>0$ and   $\frac{H}{\epsilon}$ and $\frac{\tilde{H}}{\tilde{\epsilon}}$ are positive quantities,
we have that $\Omega$ is always positive. The positiveness of $\Omega$ has a strong consequence on the fixed points in both frames. Recalling, in the Jordan frame, the fixed points are the solutions of the set of equations
\begin{equation}
\frac{dU_{i}}{dN}=f_{i}(U_j)=0,
\end{equation}
whereas the fixed points in the Einstein frame are the solutions of  
\begin{equation}
\frac{d\tilde{U}_{i}}{d\tilde{N}}=\tilde{f}_{i}(\tilde{U}_j)=0.
\end{equation}
Hence, according to (\ref{neq2}) and the positiveness of $\Omega$,
both frames have exactly the same fixed points. Moreover, the linear analysis of   fixed points involves the Jacobian matrix, whose eigenvalues reveal the dynamical nature of these particular solutions. In our case, the $ij$-th matrix element of the Jacobian $J[U_i]$ in the Jordan frame and  of the Jacobian $\tilde{J}[\tilde{U}_i]$ in the Einstein frame are $\frac{\partial f_i}{\partial U_j}$ and $\frac{\partial\tilde{f}_i}{\partial\tilde{U}_j}$, respectively. We can find the relationship between them as follows
 \begin{equation}
 \frac{\partial\tilde{f}_i}{\partial\tilde{U}_j}= \frac{\partial \Omega f_i }{\partial\tilde{U}_j} =
 \Omega \frac{\partial{f}_i}{\partial {U}_j} + f_i  \frac{\partial\Omega}{\partial {U}_j},
 \end{equation}
 and it is clear that at a fixed point we have 
 \begin{equation}
 \left[ \frac{\partial\tilde{f}_i}{\partial\tilde{U}_j}\right] = \Omega \left[\frac{\partial{f}_i}{\partial {U}_j}  \right] .
 \end{equation}
Since $\Omega$ is always positive, the signs of the eigenvalues are preserved and, consequently,
 we can conclude that the nature of the fixed points (stable, unstable or saddle) are also the same in both the frames.

The equivalence between the two frames extends far beyond the linear analysis of fixed point solutions. For instance, suppose we have an attractive domain in one of the frames, {\em i.e.}, a region of the phase-space from where no solution can escape. Such regions are typically characterized by means of a Lyapunov function \cite{Arnold}. A Lyapunov function $L(x)$ for a dynamical system $\dot x = f_1(x)$ is a smooth   positive function such that
\begin{equation}
\label{Lyap}
\dot L = \left(\nabla_xL \right)f_1 < 0 
\end{equation}
along the solutions $x(t)$. It is clear from (\ref{Lyap}) that a closed surface level around  a local minimum  of a Lyapunov function can describe an attractive domain of the phase-space since any solution, once crossing such surface, cannot return.  Repulsive domains can be defined analogously. In our case, a
Lyapunov function  $L$ in both frames  will obey
\begin{equation}
\label{Lyap1}
\frac{dL}{d\tilde N} = \Omega \frac{dL}{d  N}.
\end{equation}
Since the level surfaces in both frames are identical, the positiveness of $\Omega$ assures that attractive/repulsive domains are exactly the same in the two frames. Notice that the relation (\ref{Lyap1})
is valid for any phase-space function, it is not restricted to Lyapunov functions, and so any dynamical observable will also obey (\ref{Lyap1}) in both frames.

As already said, the equivalence will be complete, meaning a topological equivalence, if $\Omega=1$. A closer inspection of (\ref{Om1}) shows that this corresponds to $U_3=0$ and, from (\ref{U3}) we have that this corresponds the case of constant $R$ as, for instance, the case of de Sitter solutions. Our dynamical system formulation can be effectively used to determine under which conditions a general $f(R)$ theory of gravity will admit or not attractive asymptotic de Sitter solutions among other cosmological
scenarios, and these issues are now under investigation \cite{Chakraborty:2018bxh}.

\section{Final remarks}

We have considered homogeneous but anisotropic Bianchi-I universes with an anisotropic barotropic fluid
in $f(R)$ gravity in both Jordan and Einstein frames. We have shown that both frames are free from
anisotropic singularities and well defined when 
$F=f'(R)>0$ and $F'>0$ and, in this case, the introduction
of a new set of dynamical variables    allowed us to establish
a complete one-to-one correspondence between the phase-spaces in the two frames. Even though the dynamical formulation in the two frames are not topologically equivalent, we have shown that their dynamical behaviors are fully equivalent, with preserved fixed points, attraction basins and any other dynamical property which can be described as smooth functions on the phase-spaces. Our results can help not only the construction of viable   $f(R)$  cosmological models, but also contribute  to the still active debate on the physical interpretation two frames. From the dynamical point of view, if both exist, they are completely equivalent.  

Let us return to  the discussion of the invertibility of $U_2(R)$ given by (\ref{U2}), which was implicitly
used in the definition of $\gamma(U_2)$ in (\ref{gamma}). We can invert $U_2(R)$ provided
\begin{equation}
U_2' = \frac{\kappa fF'}{2F^3} \ne 0
\end{equation}
in the Jordan frame. It may seem that we need also to assume $f\ne 0$ in order to have a consistent formulation, but this is not really necessary. If $f$ changes its sign, we will indeed have two possible branches to invert $U_2(R)$, and this must be done
judiciously taking into account the smoothness of the solutions. However, the Einstein frame have exactly the same problem, one needs to invert $\tilde U_2(\phi)$ given by (\ref{tildeU2}), an from (\ref{VE}) we see that the situation is exactly the same. Even these intricacies of the dynamical formulation of both frames are completely equivalent.

\acknowledgements
S.C. and A.S.   thank the warm hospitality of the Yukawa Institute for Theoretical Physics at Kyoto University, Kyoto, Japan,  where this work was
initiated  during the long-term workshop YITP-T-17-02 ``Gravity and Cosmology 2018''.
A.S. is also grateful to FAPESP (grant 2013/09357-9) and CNPq for the financial support. The authors also wish to thank C. Corda, A. Karam, S. Odintsov, N. Ohta, A. Paliathanasis, M. Rinaldi, C.  Steinwachs, and D. Vernieri for valuable comments and suggestions.


\begin{references} 

\bibitem{RMP}  T.P. Sotiriou and V. Faraoni, Rev. Mod. Phys. {\bf 82},  451 (2010). [arXiv:0805.1726]

\bibitem{LR} A. de Felice and S, Tsujikawa, Living Rev. Rel.   {\bf 13} , 3 (2010). [arXiv:1002.4928]

\bibitem{Rev1} S. Nojiri and S. Odintsov, 	Phys. Rep. {\bf 505},  59 (2011). [arXiv:1011.0544]

\bibitem{Rev2} S. Nojiri, S. Odintsov, and V.K. Oikonomou, Phys. Rep. {\bf 692},  1 (2017)  [arXiv:1705.11098]

\bibitem{D1} A. Alho, S. Carloni, and C. Uggla,  JCAP {\bf 08}, 064 (2016). [arXiv:1607.05715]

\bibitem{D2} S.D. Odintsov and V.K. Oikonomou,  Phys. Rev. D {\bf 96}, 104049 (2017). [arXiv:1711.02230]

\bibitem{D3} S.D. Odintsov and V.K. Oikonomou,  Phys. Rev. D {\bf 98}, 024013 (2018). [arXiv:1806.07295]

\bibitem{D4} S.S. da Costa, F.V. Roig, J.S. Alcaniz, S. Capozziello, M. de Laurentis, M. Benetti,
Class. Quantum Grav. {\bf 35},  075013 (2018). [arXiv:1802.02572]

\bibitem{MagnanoSokolowski}G. Magnano and L.M. Sokolowski, 
Phys. Rev. D{\bf 50}, 5039 (1994). [arXiv:gr-qc/9312008]

\bibitem{ABGS} L.R. Abramo, L. Brenig, E. Gunzig, and A. Saa, 
Phys. Rev. D{\bf 67},  027301 (2003). [arXiv:gr-qc/0210069]



\bibitem{Bhattacharya:2015nda}
K. Bhattacharya and S. Chakrabarty,
 JCAP 1602(02), 030  (2016). [arXiv:1509.01835]

\bibitem{Dolgov:2003px}
A.~D. Dolgov and M. Kawasaki,
{\ Phys. Lett.}, B{\bf 573}, 1 (2003). [arXiv:astro-ph/0307285]

\bibitem{Faraoni:2006sy}
V. Faraoni,
  {  Phys. Rev.}  D{\bf 74}, 104017 (2006). [arXiv:astro-ph/0610734]
 
\bibitem{GSBFRF}  E. Gunzig, A. Saa, L. Brenig, V. Faraoni, T. M. Rocha Filho, and A. Figueiredo,
Phys.Rev. D {\bf 63}  067301 (2001). [arXiv:gr-qc/0012085]

\bibitem{ABG}  L. R. Abramo, L. Brenig, E. Gunzig, Phys. Lett. B {\bf 549}, 13  (2002).
[arXiv:gr-qc/0205022]

\bibitem{Figueiro:2009mm}
M.F. Figueiro and A. Saa, { Phys. Rev.} D{\bf 80}, 063504  (2009). [arXiv:0906.2588]

\bibitem{e1} S. Capozziello, S. Nojiri, S.D. Odintsov, and A. Troisi, Phys. Lett. B {\bf 639}, 135 (2006). [arXiv:astro-ph/0604431]
\bibitem{e2} F. Briscese, E. Elizalde, S. Nojiri, and S.D. Odintsov, 	Phys. Lett. B {\bf 646}, 105 (2007). [arXiv:hep-th/0612220]
\bibitem{e20} S. Capozziello, F. Darabi, and D. Vernieri, Mod. Phys. Lett. A {\bf 25}, 3279  (2010). [arXiv:1009.2580]
\bibitem{e21} C. Corda, Astropart. Phys. {\bf 34},  412 (2011). [arXiv:1010.2086]
\bibitem{e22} M. Tsamparlis, A. Paliathanasis, S. Basilakos, S. Capozziello,
Gen. Rel. Grav. {\bf 45}, 2003 (2013). [arXiv:1307.6694]
\bibitem{e3} S. Bahamonde, S.D. Odintsov, V.K. Oikonomou, and M. Wright, Annals Phys. {\bf 373}, 96 (2016). [arXiv:1603.05113]
\bibitem{e4} D. J. Brooker, S. D. Odintsov, and R.P. Woodard, Nucl. Phys. B {\bf 911}, 318 (2016). [arXiv:1606.05879]
\bibitem{e5} S. Bahamonde, S.D. Odintsov, V.K. Oikonomou, and P.V.Tretyakov, Phys. Lett. B {\bf 766}, 225 (2017). [arXiv:1701.02381]
\bibitem{e500} A. Karam, T. Pappas, and K. Tamvakis, Phys. Rev. D {\bf 96}, 064036 (2017). [arXiv:1707.00984]
\bibitem{e501} A. Karam, A. Lykkas, and K. Tamvakis, Phys. Rev. D {\bf 97}, 124036 (2018). [arXiv:1803.04960]
\bibitem{e51} N. Dimakis, A. Giacomini, and A. Paliathanasis, 	Eur. Phys. J. C {\bf 78}, 751 (2018). [arXiv:1807.00377]
\bibitem{e52} M. Rinaldi, Eur. Phys. J. Plus  {\bf 133}, 408 (2018). [arXiv:1808.08154]

\bibitem{e61} A.Y. Kamenshchik, C.F. Steinwachs, Phys. Rev. D {\bf 91}, 084033 (2015). [arXiv:1408.5769]
\bibitem{e62} M.S. Ruf, C.F. Steinwachs, Phys. Rev. D {\bf 97}, 044050 (2018). [arXiv:1711.07486]
\bibitem{e63} N. Ohta, Prog. Th.  Exp. Phys. {\bf 2018}, 033B02 (2018). [arXiv:1712.05175]


\bibitem{Amendola:2006we}
L. Amendola, R. Gannouji, D. Polarski, and S. Tsujikawa,
{Phys. Rev.} D{\bf 75}, 083504 (2007). [arXiv:gr-qc/0612180]

\bibitem{Carloni:2007br}
S.~Carloni, A.~Troisi, and P.~K.~S. Dunsby, {  Gen. Rel. Grav.}, {\bf 41}, 1757 (2009). [arXiv:0706.0452]

\bibitem{Carloni:2004kp}
S. Carloni, P. K.~S. Dunsby, S. Capozziello, and A. Troisi,
{ Class. Quant. Grav.}, {\bf 22}, 4839 (2005). [arXiv:gr-qc/0410046]

 

\bibitem{Carloni:2015jla}
S. Carloni, {  JCAP}  1509(09), 013 (2015). [arXiv:1505.06015]

\bibitem{Leach:2006br}
J. A. Leach, S. Carloni, and P. K.~S. Dunsby,
{ Class. Quant. Grav.}, {\bf 23}, 4915 (2006). [arXiv:gr-qc/0603012]

\bibitem{Goheer:2007wu}
N. Goheer, J. A. Leach, and P. K.~S. Dunsby, {  Class. Quant. Grav.}, {\bf 24}, 5689  (2007). [arXiv:0710.0814]

\bibitem{t1}T.S. Pereira, C. Pitrou, J.P. Uzan, JCAP {\bf 09}, 006 (2007). [arXiv:0707.0736]

 
\bibitem{EoS}O. Akarsu and C.B. Kilinc, Astr. Sp. Science, {\bf 326}, 315  (2010).
[arXiv:1001.0550]
 

\bibitem{Bhattacharya:2017cbn}
K. Bhattacharya and S. Chakraborty, 
  {\em Nonlinear anisotropy growth in Bianchi-I spacetime in metric $f(R)$
  cosmology}, [arXiv:1710.07906]
 

\bibitem{Chakraborty:2018thg}
S. Chakraborty, {  Phys. Rev.}, D{\bf 98}, 024009  (2018). [arXiv:1803.01594]

\bibitem{Paul:2014cxa}
N. Paul, S. N. Chakrabarty, and K. Bhattacharya,
 { JCAP}  1410(10), 009 (2014). [arXiv:1405.0139]
 
 \bibitem{Arnold} V.I. Arnold, {\em 
Geometrical Methods in the Theory of Ordinary Differential Equations}, Springer (1988).
 

\bibitem{Chakraborty:2018bxh}
S. Chakraborty, K. Bamba, and A. Saa, 
  {\em Dynamical systems approach to Bianchi-I spacetimes in $f(R)$
  gravity}, to appear.




\end{references}
\end{document}